\documentclass[12pt]{iopart}

\usepackage{graphicx,latexsym,color}

\def\be{\begin{equation}}

\def\ee{\end{equation}}

\def\bea{\begin{eqnarray}}

\def\eea{\end{eqnarray}}

\def\bi{\begin{itemize}}

\def\ei{\end{itemize}}

\begin{document}

\title[Damped Bloch Oscillations]{Damped Bloch Oscillations of Bose-Einstein Condensates in Disordered Potential Gradients}

\author{S.~Drenkelforth$^1$, G.~Kleine~B\"uning$^1$, J.~Will$^1$, T.~Schulte$^{1}$, N.~Murray$^1$, W.~Ertmer$^1$, L.~Santos$^{2}$, and J.J.~Arlt$^1$}

\address{$^1$ Institut f\"ur Quantenoptik, Leibniz Universit\"at Hannover, Welfengarten~1, D-30167~Hannover, Germany}
\address{$^2$ Institut f\"ur Theoretische Physik, Leibniz Universit\"at Hannover, Appelstra\ss e~2, D-30167~Hannover, Germany}
\ead{drenkelforth@iqo.uni-hannover.de}

\begin{abstract}
We investigate both experimentally and theoretically disorder induced damping of Bloch oscillations of Bose-Einstein condensates in optical lattices. The spatially inhomogeneous force responsible for the damping is realised by a combination of a disordered optical and a magnetic gradient potential. We show that the inhomogeneity of this force results in a broadening of the quasimomentum spectrum, which in turn causes damping of the centre-of-mass oscillation. We quantitatively compare the obtained damping rates to the simulations using the Gross-Pitaevskii equation. Our results are relevant for high precision experiments on very small forces, which require the observation of a large number of oscillation cycles.
\end{abstract}

\maketitle

\section{Introduction}

\begin{sloppypar}
The ability to realise ultracold quantum gases in periodic and disordered potentials has enabled detailed studies of fascinating effects originating in solid state physics. Ongoing investigations of single particle phenomena such as Anderson localisation as well as many particle effects like the Bose-Glass phase and the Mott insulator \cite{Bogdan,SchulteAnderson,SanchezPalenciaAnderson,Lugan, BoseGlass, Greiner} show the variety of possibilities these ensembles offer.  Especially the non-intuitive dynamics of quantum gases in periodic potentials is of interest for theoretical and experimental investigations, since quantum gases have enabled the first direct observation of Bloch oscillations~\cite{BlochZener} in tilted periodic potentials~\cite{Dahan,Wilkinson,Morsch}. In these systems the periodicity leads to an oscillatory motion instead of a linear acceleration of the particles subjected to an external force.

In solid state systems scattering at imperfections of the crystal structure leads to damping of Bloch oscillations on timescales much shorter
than the oscillation period itself. Therefore Bloch oscillations of electrons are only observable in semiconductor super lattices~\cite{Waschke}, where
the large spatial period leads to high oscillation frequencies, which are faster than the damping. Optical lattices on the other hand constitute perfect optical crystals and allow for the observation of long lived Bloch oscillations~\cite{Roati, Fattori, Ferrari, Gustavsson}. The experimental control of lattice parameters such as lattice depth and spacing, the possibility to detect the atomic cloud with absorption imaging and the very small momentum spread of Bose-Einstein condensates (BEC) have enabled detailed studies of this quantum effect.

A comparison of these systems gives rise to the question how the controlled addition of disorder to an optical lattice will affect the dynamics of particles in such a periodic potential. Disorder can be realised with additional optical potentials~\cite{Speckle,Lye,ClementFort,SchulteAnderson,Chen,dainty,superlattices,BoseGlass,Modug}, impurity atoms~\cite{Castin,Hamburg} or the roughness of the trapping potential close to the surface of atom chips~\cite{chiprough}. The simultaneous application of a homogeneous force and the optical disorder potential constitutes a spatially inhomogeneous acceleration.
This inhomogeneity can have important consequences for the application of Bloch oscillations as a sensitive tool for high precision measurements of small forces~\cite{Ferrari,Clade,Carusotto,Fattori,Roati}, since it leads to a dephasing of the quasimomentum and thus to a damping of the centre-of-mass oscillation~\cite{SchulteBloch}. Therefore a detailed quantitative understanding of the effect of the disorder and the underlying mechanism is indispensable for future applications.

We investigate the effect of a small disordered potential on Bloch oscillations of Bose-Einstein condensates in a 1D optical lattice potential. Our results indeed show that the inhomogeneity leads to significant damping of the centre-of-mass motion. The damping rate increases with disorder depth and is quantitatively compared to numerical simulations using the Gross-Pitaevskii equation (GPE). Furthermore we show that the disorder induced broadening of the quasimomentum distribution, which is the underlying mechanism for the damping of Bloch oscillations, reduces the fraction of atoms in the BEC.

\section{Bloch oscillations in periodic potentials}

The acceleration of particles in periodic potentials leads to an oscillatory motion instead of a linear increase in velocity. Since this is a pure single particle effect, it is possible to describe the underlying physics in a 1D model, while a quantitative analysis of the effects of disorder and interactions   require a 3D description, as reported in a previous work~\cite{SchulteBloch}. We briefly discuss the main features necessary for the understanding of Bloch oscillations while a comprehensive review can be found in~\cite{CristianiMorschArimondo}.

The periodicity of the potential implies that the eigenfunctions obey the same translational symmetry as the potential
\begin{equation}
\phi(z+d)=e^{iqd}\phi(z), \label{Translation}
\end{equation}
where the phase difference $q$ from site-to-site is called quasimomentum and $d$ is the lattice constant.   Since the eigenfunctions are periodic, it is possible to restrict the description of the dynamics to the first Brillouin zone [-$\frac{\pi}{d}$,$\frac{\pi}{d}$]. According to the acceleration theorem $q(t)=q(0)+\frac{F}{\hbar}\;t$ an additional potential gradient will cause the quasimomentum to evolve linearly in time,
because the energy offset from site-to-site leads to a linear increase of the phase difference from site-to-site during the time evolution. In combination with the periodicity of the band structure this causes an oscillatory motion, since the group velocity is proportional to the derivative of the band structure. The oscillation period $T=2\,\pi\,\hbar / F\,d$  depends on  the applied force $F$ and the lattice constant. The resulting amplitude $z_{BO}=\Delta / 2|F|$ is given by the width of the first band $\Delta$ and the force. Since this amplitude is only a few   micrometres for typical experimental parameters, the Bloch oscillations are conveniently analysed in momentum space via time-of-flight (TOF) absorption imaging.

While the oscillation can be described in a simple 1D model, the complex dynamics in a disordered potential gradient and the
role of interactions have to be analysed with a full 3D model. All numerical simulations presented in this work are obtained using the 3D Gross-Pitaevskii equation
\begin{equation}
\left[ -\frac{\hbar^{2}}{2m} \nabla^{2}+V_{L}(z)+V_{MF}(\textbf{r})+V_{grad}(z)+ g \left|\Psi(\textbf{r})\right|^{2}\right]\Psi(\textbf{r}) =\mu\Psi(\textbf{r}).
\label{GPE}
\end{equation}
The cylindrically symmetric magnetic trapping potential is  $V_{MF}\left(\textbf{r}\right)= \frac{1}{2} \,m\,\omega_{\rho}^{2}\, \rho^{2} + \frac{1}{2}\,m\, \omega_{z}^{2} z^{2}$ with trapping frequencies $\omega_{\rho}$ and $\omega_{z}.$ The optical potential is $ V_{L}(z)= s\, E_{r}\, \cos^{2}\left(k\,z\right)$, where $s$ is the lattice depth in units of the recoil energy $E_{r}=\hbar^{2} k^{2} / 2m$ and $k=2\pi   / \lambda$ is the wave vector of the optical lattice. The additional homogeneous gradient potential is given by $V_{grad}(z)=F \,z.$
\begin{figure}[t]
\centering
\includegraphics[width=12cm]{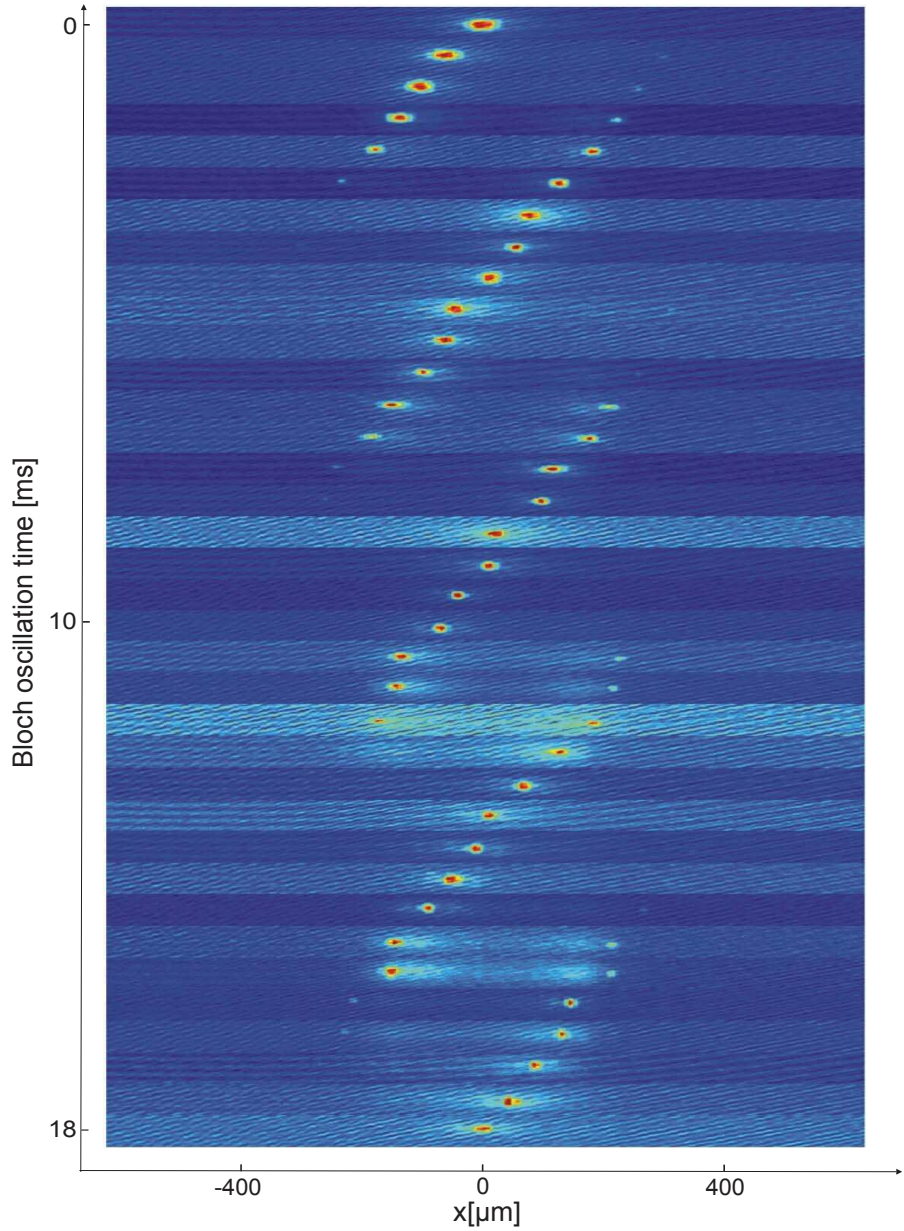}
\caption{Absorption images after a TOF of $30{\rm ms}$, for varying Bloch oscillation times. The lattice depth was $2 \, E_{r}$ and the acceleration $2.4{\rm m/s^2}$.}
\label{Absorptionimages}
\end{figure}
\begin{figure}[t]
\centering
\includegraphics[width=12cm]{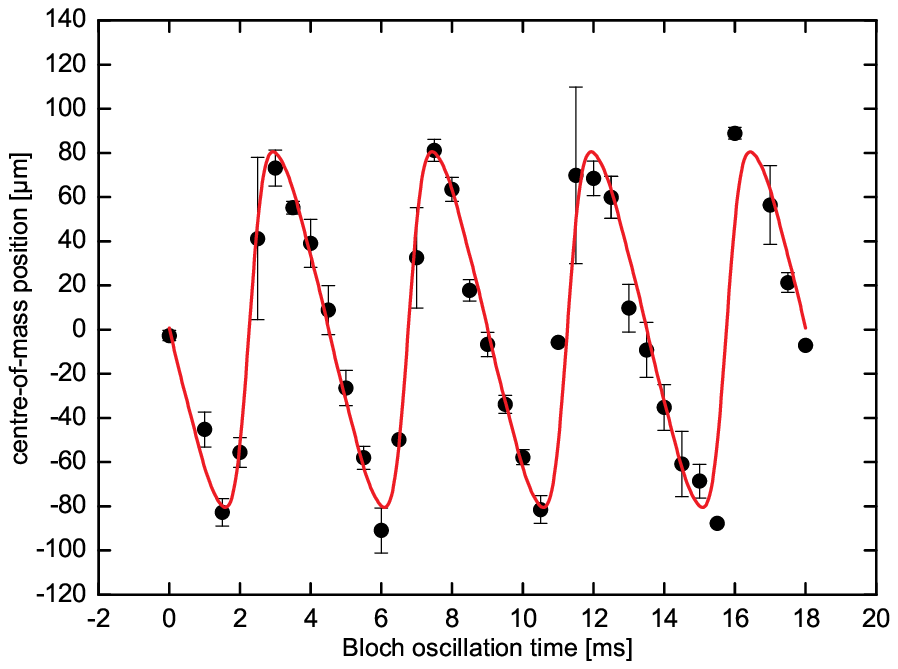}
\caption{Centre-of-mass position of Bose-Einstein condensates after a TOF of $30{\rm ms}$, as a function of the Bloch oscillation time. The parameters were identical to Fig. \ref{Absorptionimages}.}
\label{Centerofmass}
\end{figure}
\end{sloppypar}

Due to interactions between the atoms, Bloch oscillations of
Bose-Einstein condensates suffer, contrary to the case of
fermions~\cite{Roati}, from dynamical
instabilities~\cite{MenottiSupdyn,ModugnoInst}. In the outer half of
the Brillouin zone the nonlinear coupling leads to an exponential
growth of small perturbations. Independent of inhomogeneities in the
potential they are responsible for a damping of the Bloch
oscillations. Therefore it is necessary to reduce the dynamical
instability in order to investigate the effect of the disorder on
the Bloch oscillations. This can be accomplished with a combination
of a high potential gradient~\cite{cristiani} and a reduction of the
nonlinear     interactions. In recent work long lived Bloch
oscillations of Bose-Einstein condensates were
realised~\cite{Gustavsson,Fattori} by decreasing the s-wave
scattering length with a Feshbach resonance, essentially turning off
the atomic interaction. In the experiments reported here the
interactions were reduced by decreasing the density of the BEC, to
enable the observations of disorder induced damping of Bloch
oscillations.

\section{Experimental realisation}

The experiments were performed with $^{87}\rm{Rb}$ Bose-Einstein
condensates in the $\rm{F}=2, \, \rm{m}_{\rm{F}}=2$ state. A
detailed description of our apparatus is given in~\cite{Luigi}.
Nearly pure BEC of up to $N=3 \times 10^{5}$ Atoms are produced,
however all experiments described here were carried out with $N=5
\times 10^{4}$ atoms to reduce the interaction energy, and to ensure
that no significant thermal background is present. After production
of the BEC the magnetic offset field was adiabatically increased to
91G within $440{\rm ms}$, thus lowering the radial trapping
frequency to $\omega_{\bot}=2 \pi \cdot 29{\rm Hz}$. This reduces
the interaction energy by a factor of 6.25, which in combination
with low atom numbers sufficiently inhibits the dynamical
instability, such that up to four undamped Bloch oscillation periods
can be observed. The following procedure was used to observe the
Bloch oscillations. After decreasing the radial trapping frequency
the intensity of the optical lattice was adiabatically increased to
its final value within $60{\rm ms}$. Subsequently the atoms were
subjected to either a homogeneous potential gradient or the
spatially inhomogeneous one for a variable time. Finally all
potentials were turned off at the same time and the atomic cloud was
detected after a time-of-flight of $30{\rm ms}$ by absorption
imaging.

The 1D optical lattice is provided by a standing light field at a
wavelength of $825 {\rm nm}$, which is superimposed on the axial
direction of the magnetic trap with a waist of
$\omega_{0}=140\mu{\rm m}$ at the position of the atoms. The
investigations were performed at a lattice depth of $2 \,E_{r}$,
since low lattice depths lead to a large width of the energy band
and therefore to a high maximal group velocity. This results in an
oscillation amplitude of the centre-of-mass motion of $80\mu{\rm m}$
after a TOF of $30{\rm ms}$, which can easily be detected, while the
Bloch oscillations are not affected by Landau-Zener tunnelling
\cite{LandauZener}.

The homogeneous potential gradient is provided by magnetic coils in
anti-Helmholtz configuration, which produce gradients of up to
$3.7{\rm G/cm}$. This corresponds to an acceleration of $2.4{\rm
m/s^{2}}$.

The inhomogeneity is realised by a disordered optical dipole
potential, generated by imaging a randomly structured chrome
substrate radially onto the BEC, as described in a previous
publication~\cite{SchulteAnderson}. The correlation length of the
disorder is 8$\mu {\rm m}$ and its depth was varied between $0$ and
$135 \times 10^{-3} \, E_{r}$, where the depth is defined as twice
the standard deviation analogue to~\cite{Lye}.

\begin{figure}[t]
\centering
\includegraphics[width=12cm]{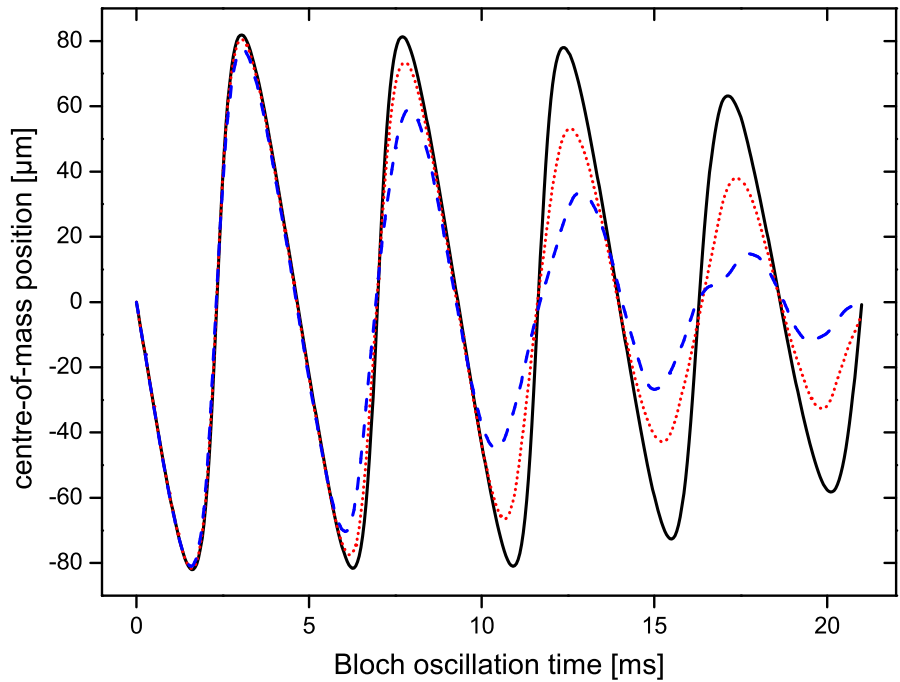}
\caption{Centre-of-mass oscillation position of a BEC for a lattice
depth of $2 \, E_{r}$, an acceleration of $2.4{\rm m/s^2}$ and
disorder depths of $0 \, E_{r}$ (black, solid line), $70 \times
10^{-3} \, E_{r}$ (red, dotted line) and $130 \times 10^{-3} \,
E_{r}$ (blue, dashed line), obtained from a numerical solution of
the GPE.} \label{DampingSimulation}
\end{figure}

Figure~\ref{Absorptionimages} shows absorption images of
Bose-Einstein condensates undergoing Bloch oscillations after
$30{\rm ms}$ TOF for a lattice depth of $2 \, E_{r}$ without
disorder. One clearly recognises the oscillation as a motion of the
central peak and the periodic appearance of a second peak replacing
the main one during an oscillation cycle. Figure~\ref{Centerofmass}
shows the centre-of-mass oscillation as well as a theoretical
prediction based on a band structure calculation without free
parameters. The position of the Bose-Einstein condensate $z_{\rm
BO}$ was calculated by numerically summing the weighted axial
positions $z_{\rm BO}=\sum_{z} \frac{N(z)}{N_{total}}z$. The
measured Bloch period was $4.5{\rm ms}$ and the oscillation
amplitude was $73\mu{\rm m}$ which is in good agreement with the
theoretical calculations of $4.66{\rm ms}$ and $80\mu{\rm m}$ for an
acceleration of $2.4{\rm m/s^2}$ and a lattice depth of $2\, E_{r}$.

\section{Damped Bloch oscillations}
\begin{figure}[th]
\centering
\includegraphics[width=10cm]{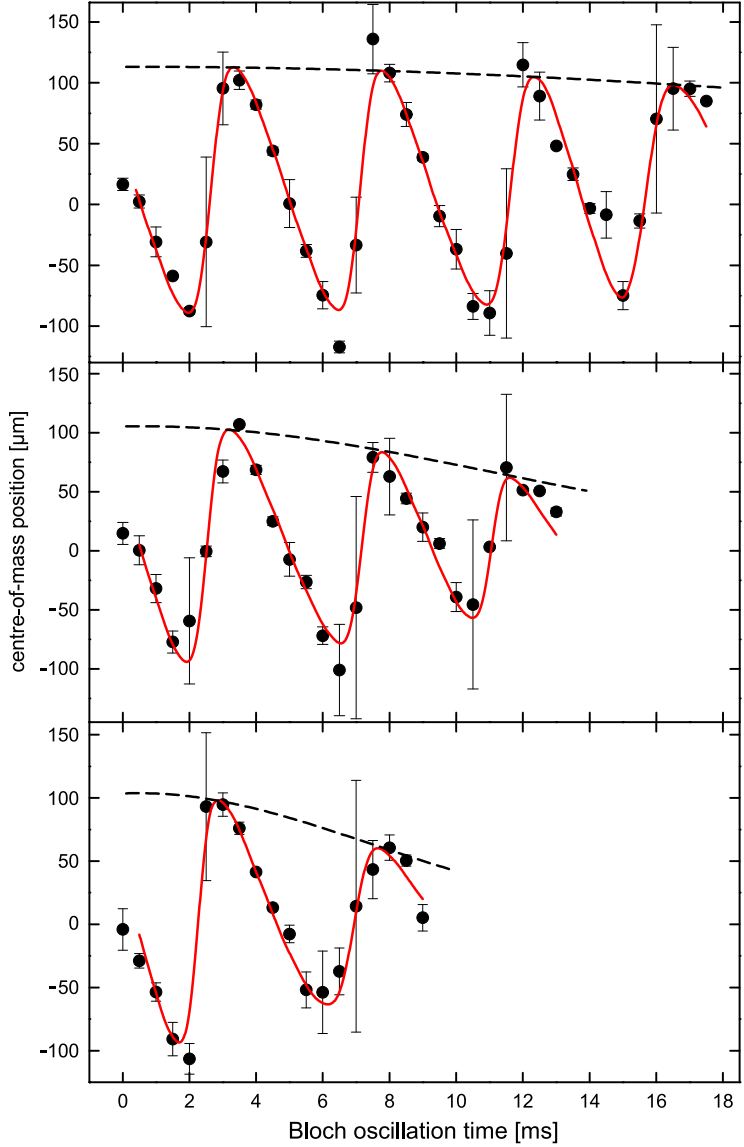}
\caption{Centre-of-mass position of Bose-Einstein condensates
performing damped Bloch oscillations. The disorder depths were $35$,
$105$ and $135 \times 10^{-3} \, E_{r}$, from top down. A fit to the
data (solid red line) and the Gaussian envelope (dashed black line)
due to the damping are also shown. Note that the recorded
oscillation time is reduced for increased disorder depth since the
broadening of the quasimomentum spectrum causes a strong reduction
of the contrast (see Fig. \ref{ContrastExperiment}). This
significantly reduces the signal-to-noise ratio of the absorption
images. } \label{dampingcenterofmass}
\end{figure}

The theory of disorder induced damping of Bloch oscillations was
investigated in detail in a previous publication~\cite{SchulteBloch}
and we only briefly review the important features for the
interpretation of the experimental results.

For the analysis of Bloch oscillations in an inhomogeneous potential
gradient the GPE (equation \ref{GPE}) has to be modified by
expanding the acceleration term to
\begin{equation}
V_{grad}=F \,z+V_{dis}(z), \label{disorderedgradient}
\end{equation}
where $V_{dis}(z)$ denotes the additional optical potential which constitutes the disorder.

Figure~\ref{DampingSimulation} shows the centre-of-mass position
obtained from numerical simulations of damped Bloch oscillations for
a typical disordered potential used in the experiment. One clearly
recognises that the damping of the oscillations strongly depends on
the disorder depth. The damping can be understood qualitatively in
terms of the evolution of the phase difference from site-to-site.
For an undisturbed homogeneous potential gradient it develops in
time according to
\begin{equation}
\Delta \phi (t)=\frac{\delta E}{\hbar}\,t,
\end{equation}
where $\delta E$ is the energy offset and $\Delta \phi$ the time dependent phase difference between neighbouring sites.
\begin{figure}[t]
\centering
\includegraphics[width=12cm]{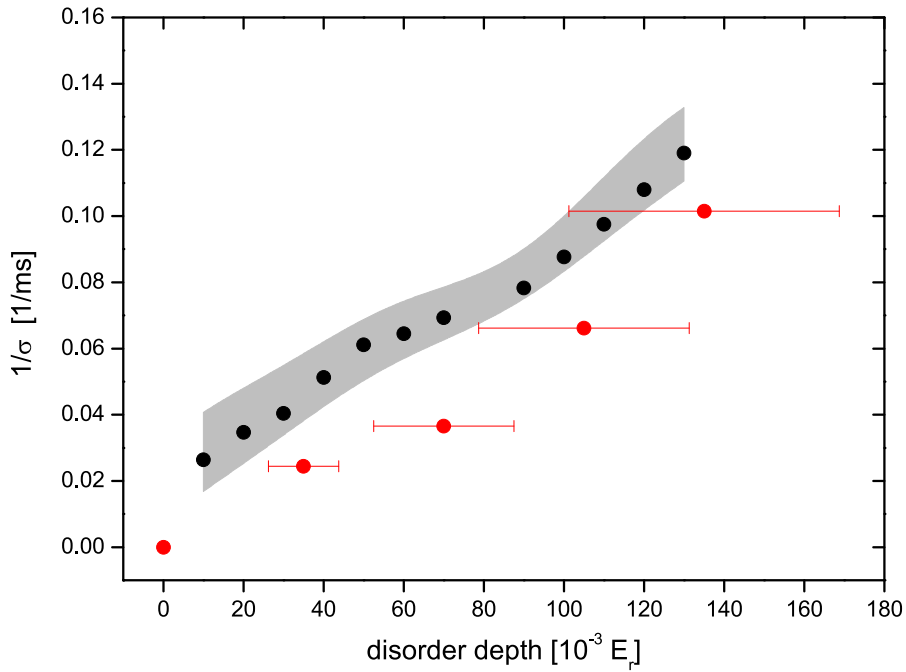}
\caption{Damping coefficient $1/ \sigma$ of the centre-of-mass
oscillation in the disordered lattice potential. The damping
coefficients were obtained by applying the same fit procedure to the
experimental data of Fig.~\ref{dampingcenterofmass} and the results
of the numerical simulations. The red dots represent the
experimental data and the black dots the simulations. Similar to the
experimental case, the number of Bloch oscillation periods used in
the fits to the simulations was reduced for increasing disorder
depth. The shaded area corresponds to atom numbers used in the
simulations ranging from  $3.5 \times 10^{4}$ to $6.5 \times
10^{4}$. } \label{damingvsdisorder}
\end{figure}

The disorder gives rise to a spatially varying energy difference
from site-to-site $\delta E(z)$. Therefore the phase evolution and
the quasimomentum vary across the lattice. This broadening of the
quasimomentum spectrum causes the damping of the Bloch oscillations,
since each $q$ corresponds to one group velocity.

\begin{figure}[t]
\centering
\includegraphics[width=8cm]{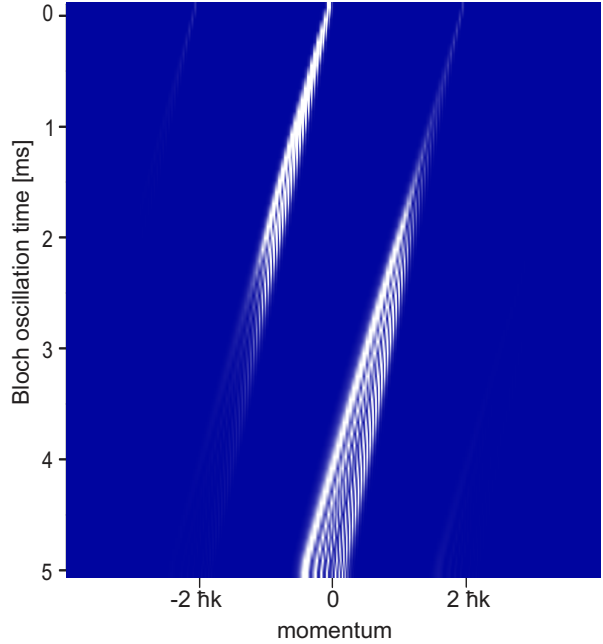}
\caption{Grey scale plot of the evolution of the momentum
distribution during a Bloch oscillation for a disorder depth of $105
\times 10^{-3} E_{r}$. The lattice depth and the acceleration were
identical to the case of Fig. \ref{DampingSimulation}.}
\label{contrastsimulation}
\end{figure}

\begin{figure}[t]
\centering
\includegraphics[width=12cm]{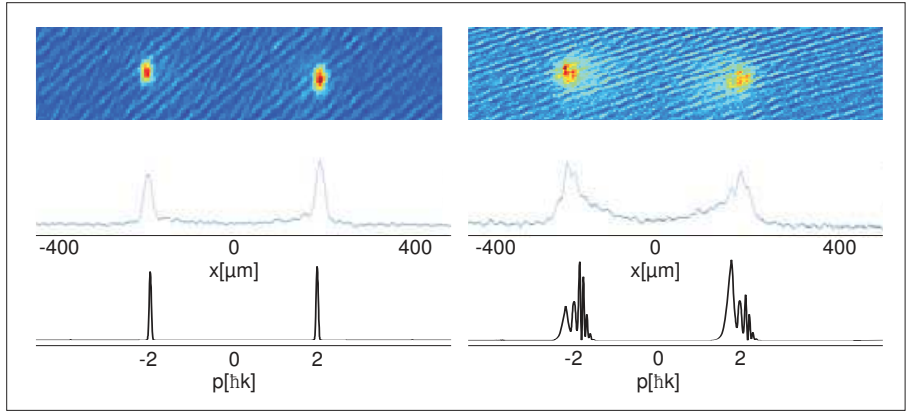}
\caption{Absorption images of Bose-Einstein condensates for an
oscillation time of $0.5$ Bloch periods at an acceleration of
$2.4{\rm m/s^2}$ and a lattice depth of $2 \, E_{r}$ after a TOF of
$30{\rm ms}$. The left column shows the case without added disorder,
whereas the right column corresponds to a disorder depth of $105
\times 10^{-3} \, E_{r}$. The top row shows the absorption images,
the middle row corresponds to the associated axial density profiles
and the bottom row contains the momentum spectrum obtained from our
simulation.} \label{bluringofbecT+E}
\end{figure}

Figure~\ref{dampingcenterofmass} shows the experimental observation
of disorder induced damping of Bloch oscillations for various
disorder depths. Note that the lattice depth and the acceleration
are identical to the undamped oscillation in
Fig.~\ref{Centerofmass}. The graph clearly shows the distinct
reduction of the oscillation amplitude for increased disorder. The
solid lines are a fit to the data, with the damping coefficient and
the periodicity as free parameters. We generate the fit function for
the damped Bloch oscillation $z_{DBO}(t)$ by multiplying the time
evolution of the undamped oscillation $z_{BO}(t)$, obtained from the
band structure calculation, with a Gaussian envelope

\begin{equation}
z_{DBO}(t)=A_0 \ e^{-t^2 / \sigma^2} \ z_{BO}(t).
\label{fitfunktion}
\end{equation}

The shape of the envelope is a consequence of the broadening of the
quasimomentum distribution in time. The width of this distribution
determines the  oscillation amplitude, which is reduced compared to
the single particle picture, because separate parts of the ensemble
simultaneously undergo different phases of the oscillation. Based on
the assumption that the width of this distribution increases
linearly in time, a Gaussian envelope of the damping amplitude is
expected~\cite{Hartmann2004,Witthaut2005}.

\begin{figure}[t]
\centering
\includegraphics[width=12cm]{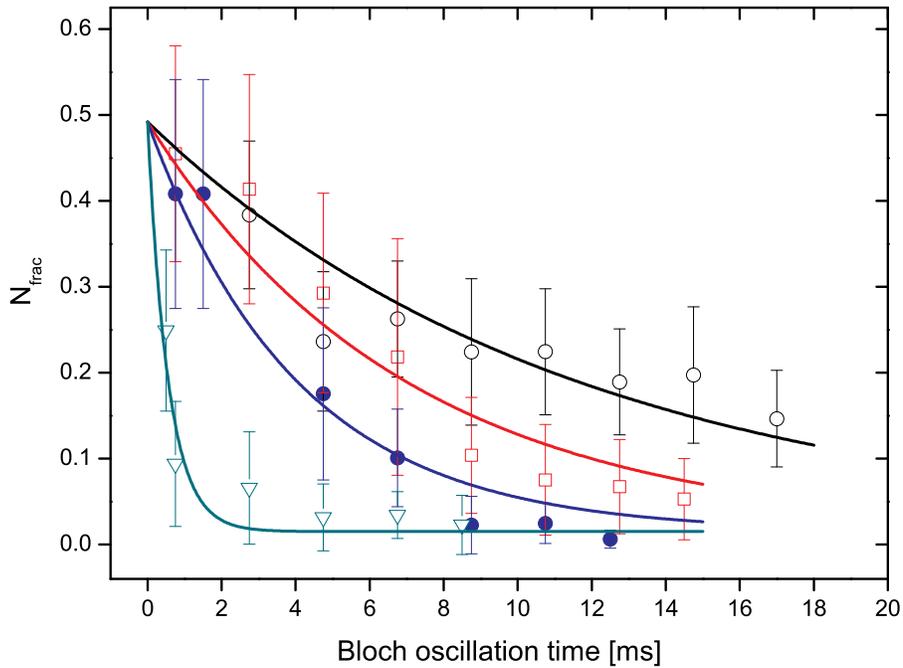}
\caption{Fraction of atoms in the BEC after a time-of-flight of 30ms
as a function of the Bloch oscillation time. Shown are four depths
of the disorder potential of $0 E_{r}$ (black circles), $35 \times
10^{-3} E_{r}$ (red squares), $70 \times 10^{-3} E_{r}$ (blue dots),
and $135 \times 10^{-3} E_{r}$ (green triangles). The curves
correspond to a fit with an exponential decay. Residual imaging
effects (see figure \ref{bluringofbecT+E}) result in an overestimate
of the total number of atoms. Therefore $N_{frac}$ is 0.5 even for
pure BEC without any discernible thermal fraction.}
\label{ContrastExperiment}
\end{figure}

\begin{figure}[t]
\centering
\includegraphics[width=12cm]{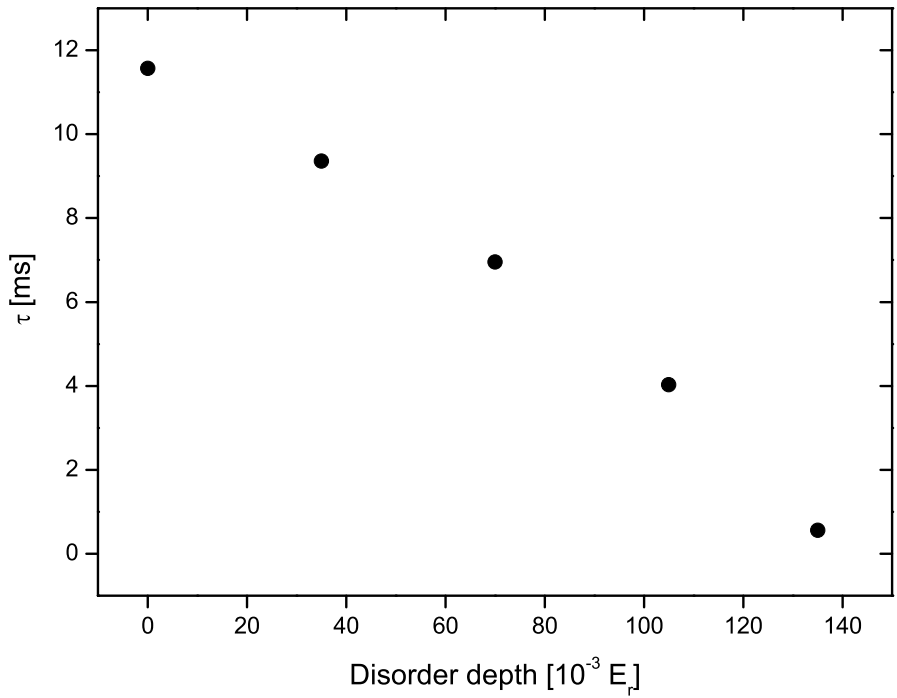}
\caption{The decay times $\tau$ of Fig. \ref{ContrastExperiment} are
shown as a function of the disorder depth. Strong reduction is
observed even for moderate disorder depths.
}\label{ContrastVersusDisorder}
\end{figure}

To quantitatively compare the experimental data to the numerical
solutions of the GPE, we show the resulting damping coefficients as
a function of the disorder depth in Fig.~\ref{damingvsdisorder}. The
parameters of the simulations correspond to the experimental ones
for a typical realisation of the disordered potential. Since the
damping rate strongly depends on the depth of the disorder
potential, the horizontal error bars represent an experimental
uncertainty in this depth of 25\%. This was estimated by evaluating
the depth of the used disorder potential at different positions,
while small deviations between the exact shape of the disorder
potential in the experiment and in the numerical simulation were not
accounted for. The shaded area corresponds to an uncertainty in the
atom number of 30\%, accounted for in the simulations. Within this
uncertainty, we observe good agreement between experimental and
theoretical values of the damping coefficients.

To show, that the broadening of the quasimomentum spectrum is the
underlying mechanism of the damping, we analyse the experimental
data with a second approach. Since the dephasing of the quasimomenta
increases the number of atoms with varying momenta, the number of
atoms in the BEC peaks is reduced~\cite{cristiani}. This picture is
confirmed by Fig.~\ref{contrastsimulation}, which shows numerical
results for the time evolution of the momentum spectrum in the
presence of a disordered potential gradient. The disorder induced
dephasing leads to an increased blurring of the sharp momentum
distribution, as the quasimomenta undergo a non-uniform evolution.

This behaviour is consistent with the absorption images shown in
Fig.~\ref{bluringofbecT+E}. In the undisturbed case sharp BEC peaks
are visible whereas the disordered case exhibits a clearly
discernible background and a broadening of the peaks due to the
dephased quasimomenta.

To analyse this effect quantitatively, we estimate the number of
atoms in the BEC $N_{\rm{BEC}}$ at different times of the Bloch
oscillation by fitting Thomas-Fermi profiles to the characteristic
peaks of the BEC.  The fraction of atoms in the BEC is calculated by
comparing the number of atoms in these peaks with the total atom
number $N_{frac}=N_{BEC}/N_{total}$.

Figure~\ref{ContrastExperiment} shows this fraction as a function of
the Bloch oscillation time for various depths of the disorder
potential. The lines are fits to the data with an exponential decay.
A clear reduction in the occupation of the BEC peaks during the
evolution of the Bloch oscillation is visible. In all cases the
dephasing of the quasimomenta precedes the onset of the damping
shown in Fig.~\ref{dampingcenterofmass}. This is in agreement with a
Gaussian shape of the damping envelope given in
Eqn.~\ref{fitfunktion} and confirms that the dephasing of the
quasimomenta is the underlying mechanism for the damping of the
centre-of-mass motion. Hence only a significant broadening of the
quasimomentum leads to a relevant damping.

Note that even at a disorder depth of $0 E_{r}$ a reduction and
therefore a dephasing is observed due to the interparticle
interactions, while the centre-of-mass oscillation in Fig.
\ref{dampingcenterofmass} is still unaffected.

The decay times $\tau$ from Fig.~\ref{ContrastExperiment} are shown
in Fig.~\ref{ContrastVersusDisorder} as a function of the disorder
depth. The decreasing fraction of atoms in the BEC peaks again
confirms that stronger disorder leads to faster dephasing of the
momentum distribution and that this broadening of the quasimomentum
is the underlying mechanism for the damping of the Bloch
oscillations.

\section{Conclusion}

We have presented the first experimental investigation on disorder
induced damping of Bloch oscillations of Bose-Einstein condensates.
The application of an additional disorder potential during the
oscillation leads to a strong damping of the centre-of-mass motion
and to a significant reduction of the fraction of atoms in the BEC.
The observed damping rates are in good agreement with predictions
based on numerical solutions of the full Gross-Pitaevskii equation
and show that the underlying physical mechanism for the damping is
the broadening of the quasimomentum spectrum due to the spatially
varying phase evolution of the condensate.

We show that even very small disorder results in fast dephasing of
the quasimomentum and therefore damping of the Bloch oscillation.
Since the disorder presented here is equivalent to a spatially
inhomogeneous force, the results are of special interest for the
application of Bloch oscillations for high precision spectroscopy of
very small forces. To reach high precision in such experiments it is
essential to follow a large number of Bloch oscillations. This
number may be reduced if the observed force is spatially
inhomogeneous on length scales comparable to the extend of the
condensate. The good agreement between theory and experiment shows
the applicability of our method to analyse the effects of spatially
varying forces, and allows for estimates of the effect of small
inhomogeneities for future experiments.

\section{Acknowledgments}

We thank M. Lewenstein for fruitful discussions. This work was supported by the Centre for Quantum Engineering and Space-Time Research and by the Deutsche Forschungsgemeinschaft within the SFB 407, the SPP1116 and within the European Graduate College Interference and Quantum Applications.\\
\\


\begin{thebibliography}{99}

\bibitem{Bogdan} B. Damski {\it et al.}, Phys. Rev. Lett. {\bf 91}, 080403 (2003).
\bibitem{SchulteAnderson} T. Schulte {\it et al.}, Phys. Rev. Lett. {\bf 95}, 170411 (2005).
\bibitem{SanchezPalenciaAnderson} L. Sanchez-Palencia {\it et al.}, Phys. Rev. Lett. {\bf 98}, 210401 (2007).
\bibitem{Lugan} P. Lugan {\it et al.},Phys. Rev. Lett. {\bf 99}, 180402  (2007).
\bibitem{BoseGlass} L. Fallani {\it et al.}, Phys. Rev. Lett. {\bf 98}, 130404 (2007).
\bibitem{Greiner} M. Greiner {\it et al.}, Nature {\bf 415}, S39 (2002).
\bibitem{Modug} J.E. Lye {\it et al.}, Phys. Rev. A {\bf 75}, 061603(R) (2007).
\bibitem{BlochZener} F. Bloch, Z. Phys. {\bf 52}, 555 (1928); C. Zener, Proc. R. Soc. A {\bf 145}, 523 (1934).
\bibitem{Dahan} M. Ben Dahan {\it et al.}, Phys. Rev. Lett. {\bf 76}, 4508 (1996).
\bibitem{Wilkinson} S.R. Wilkinson {\it et al.}, Phys. Rev. Lett. {\bf 76}, 4512 (1996).
\bibitem{Morsch} O. Morsch {\it et al.}, Phys. Rev. Lett. {\bf 87}, 140402 (2001).
\bibitem{Waschke} C. Waschke {\it et al.}, Phys. Rev. Lett. {\bf 70}, 3319 (1993).
\bibitem{Roati} G. Roati {\it et al.}, Phys. Rev. Lett. {\bf 92}, 230402 (2004).
\bibitem{Fattori} M. Fattori {\it et al.}, Phys. Rev. Lett. {\bf 100}, 080405 (2008).
\bibitem{Ferrari} G. Ferrari {\it et al.}, Phys. Rev. Lett. {\bf 97}, 060402 (2006).
\bibitem{Gustavsson} M. Gustavsson {\it et al.}, Phys. Rev. Lett. {\bf 100}, 080404 (2008).
\bibitem{Speckle} L. Guidoni, C. Trich\'{e}, P. Verkerk, and G. Grynberg, Phys. Rev. Lett. {\bf 79}, 3363 (1997); G. Grynberg, P. Horak, and C. Mennerat-Robilliard, Europhys. Lett. {\bf 49}, 424 (2000).
\bibitem{Lye} J. E. Lye {\it et al.}, Phys. Rev. Lett. {\bf 95}, 070401 (2005).
\bibitem{ClementFort} D. Cl\'{e}ment {\it et al.}, Phys. Rev. Lett. {\bf 95}, 170409 (2005); C. Fort {\it et al.}, Phys. Rev. Lett. {\bf 95}, 170410 (2005).
\bibitem{Chen} Y. P. Chen {\it et al.},  Phys. Rev. A {\bf 77}, 033632  (2008).
\bibitem{dainty} P. Horak, J.-Y. Courtois, and G. Grynberg, Phys. Rev. A {\bf 58}, 3953 (1998).
\bibitem{superlattices} R.B. Diener {\it et al.}, Phys. Rev A {\bf64}, 033416 (2001).
\bibitem{Castin} U. Gavish and Y. Castin, Phys. Rev. Lett. {\bf 95}, 020401 (2005).
\bibitem{Hamburg} S. Ospelkaus {\it et al.}, Phys. Rev. Lett. {\bf 96}, 180403 (2006).
\bibitem{chiprough} R. Folman {\it et al.}, 
Adv. At. Molec. Opt. Phys. {\bf 48}, 263 (2002); C. Henkel, P.Kruger, R. Folman, and J. Schmiedmayer, Appl. Phys. B{\bf 76}, 173
(2003); D.-W. Wang, M. D. Lukin, and E. Demler, Phys. Rev. Lett. {\bf 92}, 076802 (2004).
\bibitem{Clade} P. Clad\'{e} {\it et al.}, Phys. Rev. Lett. {\bf 96}, 033001 (2006).
\bibitem{Carusotto} I. Carusotto {\it et al.}, Phys. Rev. Lett. {\bf 95}, 093202 (2005).
\bibitem{SchulteBloch} T. Schulte {\it et al.}, Phys. Rev. A {\bf 77}, 023610   (2008).
\bibitem{CristianiMorschArimondo}  M. Cristiani {\it et al.}, Phys. Rev A {\bf 65}, 063612 (2002).
\bibitem{Witthaut2005} D. Witthaut, M. Werder, S. Mossmann, and H. J. Korsch, Phys. Rev. E {\bf 71}, 036625 (2005).
\bibitem{MenottiSupdyn} C. Menotti, A. Smerzi, and A. Trombettoni, New. J. Phys. {\bf 5}, 112 (2003).
\bibitem{ModugnoInst} M. Modugno, C. Tozzo, and F. Dalfovo, Phys. Rev. A {\bf 70}, 043625 (2004).
\bibitem{Luigi} L. Cacciapuoti, {\it et al.}  Phys. Rev. A {\bf 68}, 053612 (2003).
\bibitem{LandauZener} G. Zener, Proc. R. Soc. London A 137, 696 (1932);
\bibitem{cristiani} M. Cristiani {\it et al.}, Optics Express {\bf 12}, 4-10 (2004).
\bibitem{Hartmann2004} T. Hartmann, F. Keck, H. J. Korsch, and S. Mossmann, New J. Phys. {\bf 6}, 2 (2004);




\end{thebibliography}
\end{document}